\pdfoutput=1
\documentclass[aps,pre,reprint,groupedaddress]{revtex4-1}
\usepackage{acronym}
\usepackage{color}
\usepackage{amsmath,amssymb}
\usepackage{graphicx}
\usepackage{footmisc}
\usepackage{hyperref}

\newcommand{\x}{{\vec x}}
\newcommand{\y}{{\vec y}}
\newcommand{\n}{{\vec n}}
\newcommand{\kk}{{\vec k}}
\newcommand{\kp}{{\vec k'}}
\newcommand{\Eth}{{E_\mathrm{th}}}
\newcommand{\Ex}{{E(\vec x)}}
\newcommand{\Ey}{{E(\vec y)}}
\newcommand{\Ek}{{E(\vec k)}}
\newcommand{\Ekp}{{E(\vec k')}}
\newcommand{\corr}{{\mathcal{Q}}}

\newcommand{\dth}{{d_\mathrm{th}}}

\DeclareMathOperator\erfc{erfc}
\graphicspath{{./FIGURE/}{./}}

\DeclareMathOperator{\EX}{\mathbb{E}}% expected value

\begin{document}
% Use the \preprint command to place your local institutional report
% number in the upper righthand corner of the title page in preprint mode.
% Multiple \preprint commands are allowed.
% Use the 'preprintnumbers' class option to override journal defaults
% to display numbers if necessary
%\preprint{}

%Title of paper
\title{Activated dynamics: an intermediate model between REM and \texorpdfstring{\boldmath{$p$}}{p}-spin}

% repeat the \author .. \affiliation  etc. as needed
% \email, \thanks, \homepage, \altaffiliation all apply to the current
% author. Explanatory text should go in the []'s, actual e-mail
% address or url should go in the {}'s for \email and \homepage.
% Please use the appropriate macro foreach each type of information

% \affiliation command applies to all authors since the last
% \affiliation command. The \affiliation command should follow the
% other information
% \affiliation can be followed by \email, \homepage, \thanks as well.
\author{Marco Baity-Jesi}
\affiliation{Department of Chemistry, Columbia University, New York, NY 10027, USA}
\email[]{mb4399@columbia.edu}
%\homepage[]{Your web page}
%\thanks{}
\author{Alexandre Achard-de Lustrac}
\affiliation{Institut de Physique Th\'{e}orique, Universit\'{e} Paris Saclay, CEA, CNRS, F-91191 Gif-sur-Yvette, France}
\author{Giulio Biroli}
\affiliation{Institut de Physique Th\'{e}orique, Universit\'{e} Paris Saclay, CEA, CNRS, F-91191 Gif-sur-Yvette, France}
\affiliation{Laboratoire de Physique Statistique, \'{E}cole Normale Sup\'{e}rieure,
PSL Research University, 24 rue Lhomond, 75005 Paris, France.}

%Collaboration name if desired (requires use of superscriptaddress
%option in \documentclass). \noaffiliation is required (may also be
%used with the \author command).
%\collaboration can be followed by \email, \homepage, \thanks as well.
%\collaboration{}
%\noaffiliation

\date{\today}

\begin{abstract}
In order to study the activated dynamics of mean-field glasses, which takes place on times of order $\exp(N)$, where $N$
is the system size, we introduce a new model, the Correlated Random Energy Model (CREM), that
allows for a smooth interpolation between the REM and the $p$-spin models. 
We study numerically and analytically the CREM in the intermediate regime between REM and $p$-spin.
We fully characterize its energy landscape, which is like a golf-course but, at variance with the REM, 
has metabasins (or holes) containing several configurations. We find that 
an effective trap-like description for the dynamics emerges, provided that one identifies metabasins in the 
CREM with configurations in the trap model.
\end{abstract}

%\maketitle must follow title, authors, abstract, \pacs, and \keywords
\maketitle

% body of paper here - Use proper section commands
% References should be done using the \cite, \ref, and \label commands

When approaching the glass transition, super-cooled liquids undergo an extraordinary slowing down \cite{cavagna:09,berthier:11}. 
This phenomenon is argued by many to be related to some kind of activated dynamics \cite{biroli:13}.

Mean-field models of glasses have been particularly influential to study the glass transition: first, they capture the first decades of the slowing down by providing a microscopic 
realization of the so-called Mode Coupling Transition (MCT) \cite{reichman:05}, as shown e.g. by 
the $p$-spin spherical model which displays a MCT transition at a certain temperature 
$T_\mathrm{d}$ \cite{castellani:05}. Second, they provide a basis for the Random First Order Transition theory \cite{wolynes:12}, 
which is one of the most prominent approaches to understand the formation of glasses and provides a sound phenomenological explanation of activated dynamics of glass-forming liquids.

In order to construct a full first-principle microscopic theory of activated 
dynamics, a promising strategy consists in understanding first activated dynamics in mean-field models and then extending
this approach to finite dimensional systems.  
Pioneering results were obtained more than a decade ago by Crisanti and Ritort 
who numerically studied  mean-field models on times which diverge exponentially with the system size $N$ 
(since barriers are extensive in mean-field models) \cite{crisanti:00,crisanti:00b}. Analytical results 
have been hampered by the non-perturbative character of the activated processes.

A successful workaround to this issue has been the introduction of the Trap model (TM). It is a further simplification of usual
mean-field glasses, which captures the main features of activation and is analytically tractable \cite{dyre:87,bouchaud:92,bouchaud:95}.
%The dynamics of the TM is well-understood, so the TM can be used as a null model to help the understanding of activation in 
%more complicated glasses.
The TM is based on a drastic simplification of the dynamics: the energy landscape is envisioned as a golf-course characterized a large amount of separate minima, at the bottom of which
the system spends most of the time. Transitions between these local minima require reaching
a high threshold energy $\Eth=0$ with an Arrhenius rate of order $e^{(E-\Eth)/k_\mathrm{B}T}$.
From $\Eth$ any part of the phase space can be attained with equal
probability and in negligible time. This leads to simple and appealing description of the dynamics. 
For instance, the aging following a quench from high temperature can be precisely described in terms of an exploration of deeper
and deeper minima: the system spends a substantial fraction of the time, $t$, elapsed after the quench, in the deepest minimum visited along the dynamics. 
This leads to a logarithmic decrease of the average energy, $E(t)\simeq-T\log(t)$ \cite{bouchaud:92,bouchaud:95}.

Even though traits of the TM were searched and found in simulations of realistic glass-formers more than a decade ago \cite{denny:03,heuer:05}, only very recently it has been shown that in the limit of large-enough times and system sizes
the activated dynamics of the simplest mean-field model of the glass transition, the Random Energy Model (REM) \cite{derrida:80}, is fully TM-like \cite{gayrard:18,baityjesi:18}.
On the analytical side, this result was obtained in \cite{gayrard:18} after almost two decades of efforts on simplified versions of the REM \cite{dyre:95,benarous:02,benarous:08,cerny,gayrard:16}.
On the numerical side \cite{baityjesi:18} (see also \cite{junier:04}) this was done by extracting information on the basins through the time series of the energy, as proposed in \cite{cammarota:15}.\\
The REM is thought to represent a limiting case ($p\to\infty$) of the $p$-spin model \cite{derrida:80,gross:84}, which is
one of the most studied mean-field models of glasses \cite{castellani:05,cavagna:09,berthier:11}.
A natural question is therefore whether the TM-like description of the activated dynamics holds for this richer system as well. As a first step to solve this interesting but difficult issue, in this paper we introduce and study a variation of the REM, the Correlated Random Energy Model (CREM). In the CREM, a parameter $\alpha$ controls the amount of correlations between energy levels.
By tuning the parameter $\alpha$ it is possible to interpolate smoothly and exactly between the REM and the $p$-spin model (as discussed below, the common lore that the p-spin model tends to the REM for $p\to\infty$ is problematic when studying activated dynamics, this is the reason why we introduce the CREM). 
We study the CREM in the intermediate regime, where energy levels are weakly correlated, and show that 
although the energy landscape structure is richer with respect to the REM, TM-like dynamics holds in the CREM.

{\it The Random Energy Model (REM).}
In the REM \cite{derrida:80}, there are $N$ binary variables $s_{i}=\pm1$, called spins.
Each configuration of the $N$ spins (also called state) is assigned a random energy from
a Gaussian distribution of mean $0$ and variance $N$. The energies of different
states are independent.
Each state has $N$ neighbours, corresponding to the flipping of a single spin.
Typically, i.e. with probability one in the large $N$ limit, the energies of the neighbours are included in the interval 
$(-\sqrt{2N\log(N)},\sqrt{2N\log(N)})$, which corresponds to zero intensive energy \cite{baityjesi:18}. The majority of 
the states are also at zero intensive energy. This leads to a landscape like a golf-course where to escape an energy minimum 
the system has typically to climb up to $\Eth^\mathrm{REM}=-\sqrt{2N\log(N)}$, and configurations at low energy $E$ act like 
trap with life-time of the order $e^{(\Eth-E)/k_\mathrm{B}T}$.    
It was recently shown that in large-enough systems and on exponentially large time scales, the equilibrium and aging  
dynamics of the REM can be effectively described through the TM \cite{gayrard:18,baityjesi:18} (see also the previous 
works \cite{benarous:02,benarous:06,benarous:08,dyre:95,junier:04}).

{\it The \texorpdfstring{$p$}{p}-spin Model.}
The Hamiltonian of the $p$-spin model contains $p$-body interactions between $N$ Ising spins $s_i=\pm1$, and reads \cite{derrida:80}: 
\begin{equation}
 \mathcal{H} = -\sum_{1\leq i_1< i_2<\ldots< i_p\leq N} J_{i_1,i_2,\ldots,i_p} s_{i_1}s_{i_2}\ldots s_{i_p}\,,
\end{equation}
where the couplings $J_{i_1,i_2,\ldots,i_p}$ are extracted from a random Gaussian distribution with mean zero and variance ${\frac{p!}{2N^{p-1}}}$.\\
In the $p$-spin model, as in many other mean-field glassy systems \cite{cavagna:09}, there exists a threshold energy $\Eth$ below which dynamics 
become activated and exponentially slow (in the system size)\cite{montanari:06,benarous:17}. At variance with the REM, this threshold energy is 
extensive and negative \cite{rizzo:13}. \\
For any $p$ of order one and in the large $N$ limit, the correlation matrix between two generic configurations $\vec x=(s_1^{(x)},s_2^{(x)},\ldots,s_N^{(x)})$ and $\vec{y}=(s_1^{(y)},s_2^{(y)},\ldots,s_N^{(y)})$ 
reads \cite{derrida:80} 
\begin{equation}\label{eq:corrpspin}
\overline{E(\vec{x})E(\vec{y})} = N q(\vec x,\vec y)^p\,, 
\end{equation}
where $\overline{(\ldots)}$ is the average over different instances
of the couplings, and $q(\vec x,\vec y)=\frac1N\sum_i^N s_i^{(x)}s_i^{(y)}$ is the overlap between $\vec x$ and $\vec y$. Thus, contrary to the REM, now 
energies are correlated. Note that eq. (\ref{eq:corrpspin}) is valid at leading order in $N$ (sub-leading corrections have been neglected).
The REM can be {\it formally} recovered in 
the $p\to\infty$ limit. In fact, if one could take $p\rightarrow \infty$ before $N\rightarrow \infty$ then 
eq. (\ref{eq:corrpspin}) would lead to uncorrelated energies for different configurations. 
This limit, however, does not make sense since $p=N$ at most \footnote{For $p=N$ there is
only one coupling $J_{i_1,\ldots,i_N}$, and the system has only two states, with equal and opposite energy. }. This is not an issue for 
thermodynamics, which was indeed shown to converge to the one of the REM even for $N\rightarrow\infty$ first and $p\rightarrow \infty$ later 
\cite{derrida:80,gross:84}; it is instead an issue for activated dynamics, since uncorrelated energies are a key-ingredient for the analysis 
of the dynamics of the REM. For this reason we consider below a different model that allows 
us to interpolate continuously between the $p$-spin model with $p\sim \mathcal{O}(1)$ and the REM.
%Despite there being a well-defined threshold energy, it is believed that the $p$-spin model does not exhibit trap dynamics,
%since the energy landscape does can not be separated into well-defined separate energy basins \cite{}.

{\it The Correlated Random Energy Model (CREM).}
%Further, when $p$ becomes of order $N$ -- let us say that $p=\tilde{\alpha} N$, with $\tilde{\alpha}\lesssim1$ -- two neighboring sites
%are always correlated. In fact, one obtains that, given two neighboring configurations $\vec{x}$ and $\vec{x}'$, in the large-$N$ limit their
%correlation is $\overline{E(\vec{x})E(\vec{x}')} = e^{-2\tilde{\alpha}}$. Thus, the REM, where different states are uncorrelated, 
%can never be recovered in this setting.
We consider a variant of the REM, with correlated energies. We call this new model the Correlated Random 
Energy Model (CREM)~\footnote{The similar but distinct Generalizated REM (GREM) was introduced in the past \cite{derrida:85} with the purpose of studying 
spin-glasses. Hence, the aim of the generalization, as well as the correlations between energies (chosen to match the Parisi solution of the 
Sherrington-Kirkpatrick model \cite{parisi:80b}) are different.}. 
In the CREM, there are $N$ spins $s_i=\pm1, (i=1,\ldots,N)$, so there are $2^N$
different states. As in the REM, each state is assigned a random Gaussian energy of mean 0 and variance 1. However, in the CREM the energies of two configurations 
$\vec{x}$ and $\vec{y}$ are not independent. Their covariance is
\begin{equation}\label{eq:corrcrem}
 \overline{E(\vec{x})E(\vec{y})} = N q(\vec x, \vec y)^{\alpha N},\
\end{equation}
where $\alpha\in[0,\infty)$ is a parameter. Contrary to~\eqref{eq:corrpspin}, the equation above is strictly valid for any $\alpha$ and $N$. 
The largest covariance is obtained for nearest neighboring configurations $\vec{x}$ and $\vec{x}'$. 
Since they differ by one-spin flip only, in the large-$N$ limit one finds $\overline{E(\vec{x})E(\vec{x}')} = N e^{-2\alpha}$.  The parameter
$\alpha$ allows for a smooth interpolation between the REM and the $p$-spin model. In fact, for $\alpha=\frac{p}{N}$, Eq.~\eqref{eq:corrcrem} reduces 
to Eq.~\eqref{eq:corrpspin}, and the $p$-spin model is recovered. Whereas if $\alpha$
diverges with $N$ then the energies become independent variables at large $N$,
as in the REM. In the following, in order to study activated dynamics in an intermediate case between REM and $p$-spin, we focus on the regime $\alpha$ of the order of one. \\
The study of the CREM can be easily implemented numerically, since its computational complexity
does not increase with $p$. As a matter of fact, by going to Fourier space on the hypercube the 
energies of the CREM become independent; so one can generate them as Gaussian independent random 
variables in Fourier space, and then antitransform them back to real space (see App.~\ref{app:fourier}).

{\it A golf-course with structure in the holes.}
For $\alpha$ of order one and very large $N$, the correlation matrix reads
\begin{equation}\label{eq:corrcremintermediate}
 \corr_{xy} \equiv \overline{E(\vec{x})E(\vec{y})} = N e^{-2\alpha r_{xy}},\
\end{equation}
where $r_{xy}=\frac12\sum_{i=1}^N|s_i^{(x)}-s_i^{(y)}|$, the number of spins that are different between $\vec x$ and $\vec y$,
indicates the \emph{distance} between the two configurations. The exponential decay of $\corr_{xy}$ with the distance determines a kind of correlation length $
 \xi=\frac1{2\alpha}$ for the typical size of correlated domains on the hypercube 
\footnote{Mind that $\xi$ is a distance in phase space.}.\\
%In the $p$-spin limit the correlations are extensive, so $\xi$ would be diverging.
Since the energy distribution is Gaussian, one can easily obtain, 
given a configuration $\vec x$ with energy $E_x$, the conditional probability of a configuration $\vec y$ at distance $r$ from it:
\begin{equation}\label{eq:pe-cond}
 P(E_y|E_x) = \frac{1}{\sqrt{2\pi N(1-\rho^{2r})}}e^{-\frac{(E_y - \rho^rE_x)^2}{2N(1-\rho^{2r})}}\,,
\end{equation}
where we defined $\rho=e^{-2\alpha}$. From Eq.~\eqref{eq:pe-cond} one can get the expectation of the energy $E_y$ conditioned 
to $E_x$, $\EX[E_y|E_x] = \rho^r E_x$. Thus, we find that for $\alpha=p/N\to0$, the usual $p$-spin case, the energy landscape a 
finite number of steps away from a given configuration is flat, with every energy almost equal to $E_x$, consistently with an infinite correlation length $\xi\to\infty$. When $\alpha$ diverges, the REM case, $\EX[E_y|E_x] = 0$ and $E_y$ is independent from $E_x$, consistently
with a vanishing correlation length. For $\alpha\sim1$, one has instead an intermediate situation in which the energies of neighboring configurations are typically higher but do not reach directly zero intensive energy. The energy landscape is still golf-course like but has gained some more structure with respect to the REM.

{\it Complexity of "critical points" and threshold energy.}
In analogy with calculations on the spherical $p$-spin model \cite{crisanti:95,cavagna:98,crisanti:03b,benarous:13}, and with other systems displaying complex energy landscape \cite{wainrib:13,fyodorov:16,ros:18}, 
we focus on the discrete counterparts of critical points in continuous systems. Given a configuration $\vec x$, we call it "a critical 
point of order $k$" if exactly $k$ of its neighbors have lower energy. For example, local minima correspond to $k=0$. In the case of 
large funnel-like basins, the analogous of a saddle of order 1 connecting them, would be a configuration with $k=2$. Note, yet, 
that even though two local minima must be connected by a configuration with $k\geq2$,
it is not true that every configuration with $k\geq2$ connects different minima 
\footnote{This is due to the hypercube structure of the phase space. If instead the phase space graph was a tree
around every local minimum, then configurations with $k\geq2$ would always connect two different basins.}.
The choice of considering only nearest neighbor configurations to define "critical points" makes sense for 
the CREM and the REM where the energy changes substantially by one spin-flip.  

\begin{figure}[tb]
 \includegraphics[width=\columnwidth]{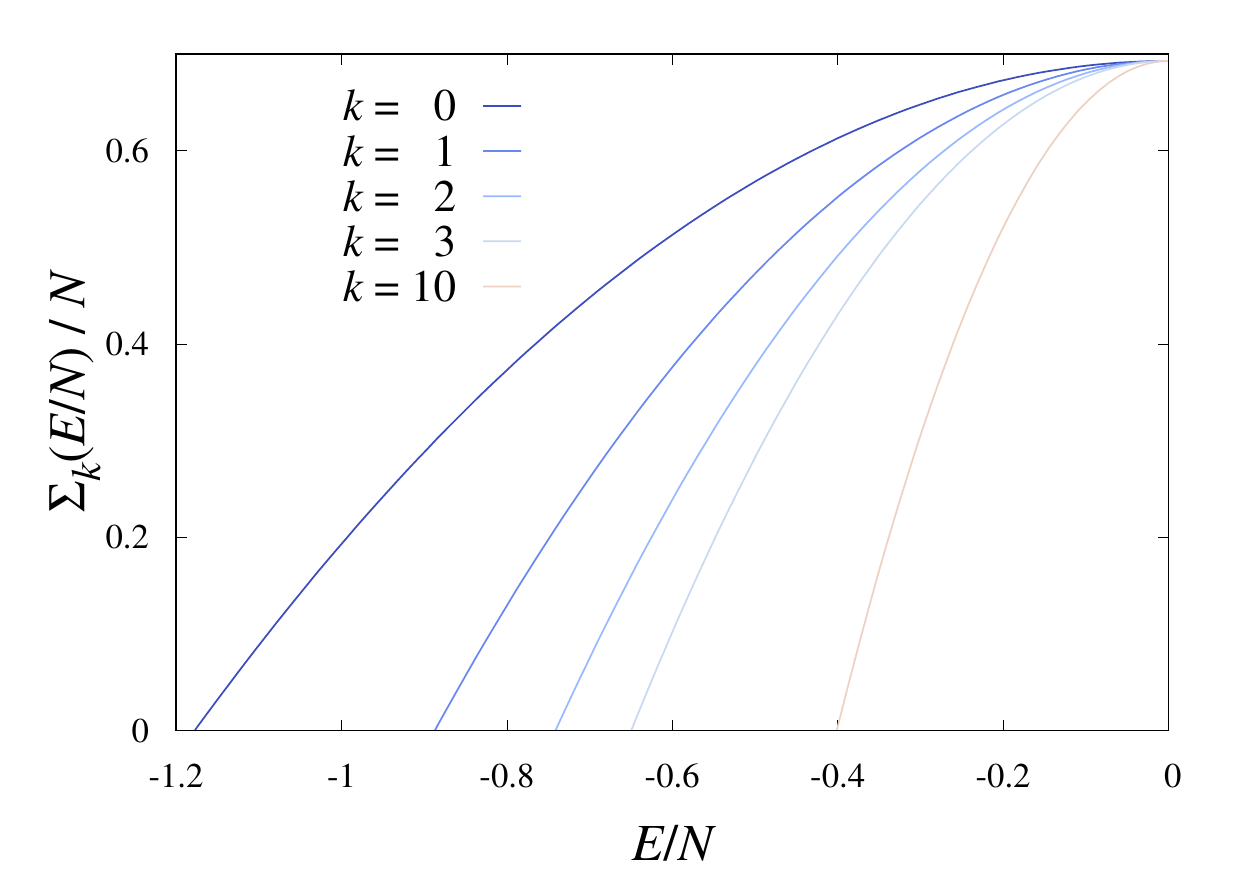}
 \caption{Complexity curves $\Sigma_k$s in the CREM, for $\alpha=1$.}
 \label{fig:complexity}
\end{figure}

The probability that a configuration with negative energy $E$ be a "critical point of order $k$" is 
\begin{equation}\label{eq:pe-sad}
 P_k(E) = \binom{N}{k} \left(  \frac{\erfc \left[B(E)\right]}{2}    \right)^{k}       
                         \left(1-\frac{\erfc \left[B(E)\right]}{2}    \right)^{N-k}
 \end{equation} 
where $B(E)=-E\sqrt{\frac{(1-\rho)}{2N(1+\rho)}}$ and $\erfc(x)$ is the complementary error function. 
As shown in App.~\ref{app:order}, this equation can be easily established by realizing that after conditioning on the 
value of the energy $E$ of a given configuration, the energies of all its nearest neighbors are independent.  
%we used Eq.~\ref{eq:pe-cond} to establish the probability of $k$ neighbors having an energy lower than $E$ and
%$N-k$ having a higher one, and $B(E)=-E\sqrt{\frac{(1-\rho)}{2N(1+\rho)}}$ diverges in the thermodynamic limit
%when intensively negative energies (i.e. $e\equiv E/N\sim-1$) are taken into account.
For large $N$, one can use Eq.~\eqref{eq:pe-sad} to calculate the entropy of "critical points of
order $k$", which we call \emph{complexity} $\Sigma_k$ in analogy with the one of the spherical 
$p$-spin \cite{crisanti:95,cavagna:98,crisanti:03b,benarous:13}. To leading order one gets (see App.~\ref{app:complexity})
\begin{equation}\label{eq:complexity}
 \frac{\Sigma_k(E)}{N} = \log(2) - \frac{E^2}{2N^2}\left( 1+ k\frac{1-\rho}{1+\rho} \right)\,.
\end{equation}
We plot $\Sigma_k$ for several $k$ in Fig.~\ref{fig:complexity}, to stress its qualitative similarity with
the complexity in the $p$-spin model (with the already-mentioned \emph{caveat} on the analogy with saddle points).
Even though present, non-minimum configurations are exponentially fewer than minima for any $E<0$, indicating that
the intensive threshold energy is zero, as in the REM but differently from the $p$-spin \cite{cavagna:09,baityjesi:18}.
This also implies that for any $E<0$, to leading order in $N$, the entropy is equal to the complexity of the minima: 
$S(E)=\Sigma_0(E)$. Since $S(E)$ is the same as in the REM \cite{derrida:80}, the thermodynamics is the same in both models (even 
though the $\Sigma_k$s for $k>0$ are different). 

In order to calculate more precisely the threshold energy, we say that a configuration of energy $E$ is at $E_\mathrm{th}$ 
if the lowest-lying among its neighbors has energy $E_\mathrm{min}^\mathrm{(neigh)}(E)=E$. The quantity 
$E_\mathrm{min}^\mathrm{(neigh)}(E)$ is found by combining Eq.~(\ref{eq:pe-cond}) with common results from extreme 
statistics (see App.~\ref{app:eth}),
resulting in
\begin{equation}\label{eq:eth}
 \Eth = -\sqrt{2\frac{1+\rho}{1-\rho}N\log(N)} = \sqrt{\frac{1+\rho}{1-\rho}}\,\, \Eth^\mathrm{REM}\,.
\end{equation}
As expected from the "complexity" calculation, even though correlations have the effect of lowering the threshold 
energy with respect to the REM, $\Eth$ is still intensively zero for large $N$.

{\it Size and structure of metabasins.}
We now focus on metabasins, defined as sets of configurations connected by paths that do not overcome $E_\mathrm{th}$. 
The threshold energy can be used to define a typical linear size $d_\mathrm{th}$ of the metabasins, as the 
typical minimum number of spin flips required, starting from a local energy minimum, to reach 
$\Eth$. In the REM case, one needs just a spin flip, i.e. the typical meta-basins consists of 
a single configuration. This is no longer the case for the CREM. 
%Note that even though there could be some local structure, with smaller basins inside a metabasin,
%the metabasin is what matters for the long-time dynamics \cite{}.
Since every configuration has $N$ neighbors, as long as $d_\mathrm{th}\ll N$, a rough estimate of the phase space volume 
$\Omega_\mathrm{b}$ of a basin is $\Omega_\mathrm{b}\sim N^{d_\mathrm{th}}$.
According to Eq.~\eqref{eq:pe-cond}, given an energy minimum $\vec x$ at energy $E_x=E$, configurations $\vec y$ at distance $r$ from it have in average an energy $E_y=\rho^rE_x$. 
The linear size $d_\mathrm{th}(E_x)$ is therefore found by imposing that $E_y$ reach the threshold level: $\rho^{d_\mathrm{th}(E_x)} E_x=\Eth$.
The resulting linear metabasin size is therefore: 
\begin{equation}
 d_\mathrm{th}(E) = \frac{1}{2\alpha}\log\left(\frac{E}{\Eth}\right)\,.
\end{equation}
From this equation one sees that deep configurations (meaning with negative intensive energy) are in metabasins made of multiple configurations, 
and that the metabasins become larger and larger as $\alpha$ decreases
\footnote{One could average $\dth(E)$ over the energy, to obtain the average basin size $\overline{\dth}$, which is of order 1 with $\mathcal{O}(\log\log N)$ corrections. 
Yet, the most relevant basins for the dynamics are the deepest ones, which are exponentially rare, so $\overline{\dth}$ is not a good estimator because it privileges
the high-energy minima that are quickly escaped.}.

To obtain further information on the structure of the metabasins, we focus on the distance $d_\mathrm{sad}(E)$ between a local minimum at 
energy $E$ and the closest configuration of order $k\geq2$.
In a continuous system, this second definition would correspond to the distance between a local minimum and the nearest saddles
Since to leading order there are $N^d$ configurations at distance $d$, the typical distance $d_\mathrm{sad}$ from the minimum 
at which one finds a saddle is obtained by imposing $N^{d_\mathrm{sad}(E)} P_k(\rho^{d_\mathrm{sad}(E)}E)\sim1$.
As we previously showed (Eq.~\eqref{eq:complexity}), the saddles of order $k=2$ are overwhelmingly more common that those of higher order, thus 
one can impose the simpler condition $N^{d_\mathrm{sad}(E)} P_2(\rho^{d_\mathrm{sad}(E)}E)\sim1$, which to leading order yields
$d_\mathrm{sad}\sim\frac1{2\alpha}\log(\frac{E}{\Eth})$. This shows that one has to climb up to 
$\Eth$ before finding a "saddle", thus implying that a metabasin contains several configurations but no additional higher local minima.
%\footnote{The reader could argue that $\dsad$ is only a lower bound to the distance needed to reach a ``saddle'', since a configuration with $k=2$ could be connected to
%2 paths leading to the same local minimum. Yet, this distance is upper bounded by $\dth$, and since $\dsad=\dth$, the argument still holds.}. 
This is in agreement with Eq.~\eqref{eq:pe-cond}, which implies that energy typically increases by going further away from a given low energy configuration.

\begin{figure}[!tb]
 \includegraphics[width=\columnwidth]{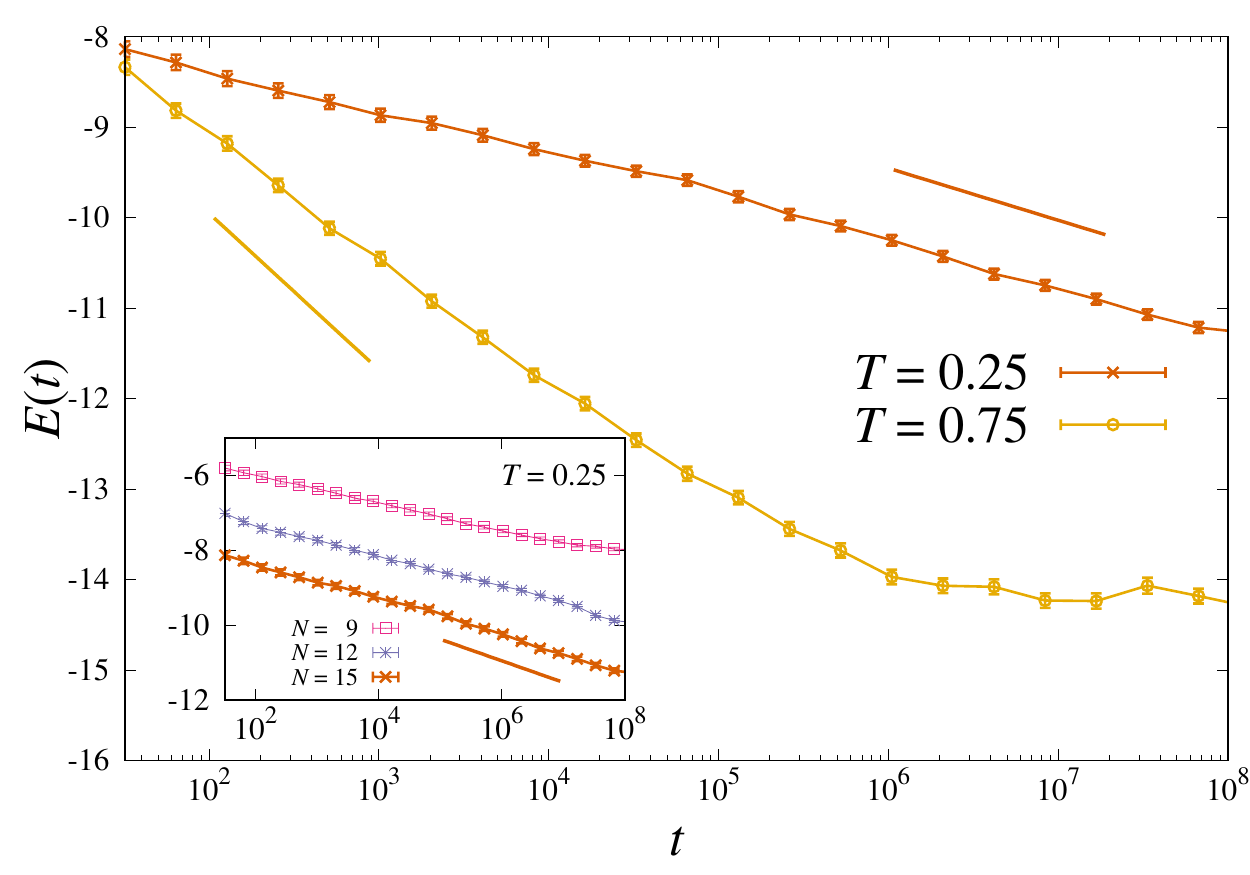}
 \caption{Energy as a function of time in the CREM in the aging regime induced by a quench from $T=\infty$ to $T=0.25$ (and $0.75$), for $\alpha=1$, $N=15$. Curves are averaged along trajectories of ~$3\times10^8$ time steps on 2250 (and 900) different instances of the disorder. The straight lines correspond to a slope $-T$. Especially at high temperature, where the equilibrium energy $\langle E(T)\rangle$ is higher, finite-size effects on the slope are expected \cite{baityjesi:18}, due to the convergence to $\langle E(T)\rangle$. The \textbf{inset} portrays the finite-size dependence of $E(t)$ for $T=0.25$.}
 \label{fig:en-t}
\end{figure}

{\it Trap dynamics.}
The previous analysis shows that when $\alpha\sim1$ the energy landscape of the CREM is more complex than the one of the REM. Nevertheless, the golf-course structure of 
the landscape still holds with the additional characteristic that the holes actually contain a large 
number of configurations. Since to escape from a hole (or metabasin) the system has to climb until 
a zero intensive threshold energy, as in the REM, an effective description in terms of trap 
dynamics should hold. The difference with the REM is that one needs to coarse grain the energy landscape: the counterpart of configurations of the trap model are metabasins in the CREM. 
As a check of Trap dynamics, we have studied the energy 
as a function of time for quenches at different temperatures. 
The resulting curves are given in Fig.~\ref{fig:en-t} for $\alpha=1$ and agree with 
the behavior $E(t)\simeq -T\log(t)$. The linear logarithmic decrease combined with the prefactor equal to the temperature are a strong indication that a description in terms of Trap dynamics holds \cite{bouchaud:92,junier:04,baityjesi:18}, as recalled in the introduction (the final plateau for $T=0.75$ is a finite size effect due to the fact that the system is small enough to eventually equilibrate).
We didn't study aging functions or distributions of trapping times since these observables present strong finite size and finite time effects, which makes 
the comparison with the trap model problematic \cite{cammarota:18} even in the REM where it is rigorously known that trap predictions hold \cite{baityjesi:18, gayrard:18}.
%Besides the long time scales involved, it is very hard to extract information through numerical simulations, because of strong finite-size effects inherent to the distribution 
%of the random energies \cite{cammarota:18}.

{\it Discussion and Conclusion.}
A central question in the study of glasses is the nature of activated dynamics. 
The dynamics of mean-field models on time-scales diverging with $N$ provides a useful and interesting paradigm. 
Yet, although some results are known \cite{crisanti:00,crisanti:00b,billoire:01}, the behavior 
of simple models such as the one with $p$-spin interactions or variants of it has not 
been fully elucidated. Only the dynamics of the REM was completely worked out. The new model we introduced and studied in this work, the  Correlated REM, 
provides a way to bridge the gap between these systems.
%, and to study analytically the $p$-spin model with discrete spins.
We studied it in an intermediate regime between REM and $p$-spin, and found that its energy landscape and its 
dynamics are trap-like provided that one identifies metabasins with single configurations of the trap model. 
The next crucial step is therefore to understand to what extent trap dynamics hold in the $p$-spin case too. A likely possibility is that during the activated 
dynamics the system does not have to climb up to $\Eth$, i.e. that a more complex structure of activated paths arises in the $p$-spin case. Nevertheless, some 
features of TM-like dynamics, such as partial equilibration at all energies above the one reached at time $t$ during aging \cite{junier:04}, could still effectively hold.  
Another aspect worth future analysis is the effective temperature description of the aging dynamics \cite{cugliandolo:93}, which has been 
 found to hold also in the activated regime of mean-field systems \cite{crisanti:03} but it is known to be violated in trap models \cite{fielding:02,sollich:03}. 
 To address these and other questions, further studies of the CREM dynamics can provide very valuable insights. 
 
% If you have acknowledgments, this puts in the proper section head.
\begin{acknowledgments}
We thank C. Cammarota for useful feedbacks on the manuscript.
We thank V. Astuti, C. Cammarota, I. Hartarsky, F.P. Landes, H. Miyazaki, V. Ros and P. Urbani for useful discussions.
This work was funded by the Simons Foundation for
the collaboration “Cracking the Glass Problem” (No. 454935 to G. Biroli and No. 454951 to D.R. Reichman), and
the ERC grant NPRGGLASS (No. 279950). M.B.-J. was partially supported
through Grant No. FIS2015-65078-C2-1-P, jointly funded by MINECO (Spain) and FEDER (European Union).
\end{acknowledgments}

% Specify following sections are appendices. Use \appendix* if there
% only one appendix.
\appendix
\section{Generating the energies of the CREM in Fourier space} \label{app:fourier}
In this section we show how to generate the energy levels of the CREM, by passing through
Fourier space, where the energy levels are independent.
\bigskip

The overlap between two configurations $\vec x$ and $\vec y$ is $q(\vec x, \vec y) = \frac1N\sum_i^N s_i^{(x)} s_i^{(y)}$,
where $s_i^{(x)}=\pm1$ are the spins of configuration $\vec x$.
The spins  can be rewritten as $s_i^{(x)} = (-1)^{x_i}$, with $x_i=0,1$.
This way, the overlap becomes $q(\vec x, \vec y) = \frac1N\sum_i^N (-1)^{x_i+y_i}$.
The $x_i$ ($i=1,\ldots,N$) are coordinates of a hypercube of size $L=2$, with periodic boundary conditions,
so $-x_i=x_i$. As a consequence, the overlap can be written as a difference of the degrees of freedom,
\begin{equation}
 q(\vec x, \vec y) = \frac1N\sum_i^N (-1)^{x_i-y_i}= q(\vec x - \vec y) \equiv q(\vec z)\,,
\end{equation}
where we defined $\vec z=\vec x-\vec y$.
\bigskip

We now want to show that random energies $E(\vec x)$, with the correlations defined in Eq.~\eqref{eq:corrcrem},
are independent in Fourier space, and calculate their wave-vector dependent variance $\overline{\Ek E(-\vec k)}$. 
Therefore, to generate the full set of energies of a sample, 
it is enough to generate the $\Ek$ as independent Gaussian random variables with variance $\overline{\Ek E(-\vec k)}$, 
and then take the antitransform to have the energies in real space,
\begin{equation}
 E(\vec x) = \frac{1}{2^N}\sum_{\kk} e^{i\kk\cdot\x}\Ek\,
\end{equation}
where the wave vectors take the form $\kk=\frac{2\pi}{L}(n_1, \ldots, n_N)$, and $n_i = 0, 1$,
so, since $L=2$, the antitransform can be simplified to
$E(\vec x) = 2^{-N}\sum_{\kk} (-1)^{\n\cdot\x}\Ek$.

Since the discrete Fourier transform of the energies reads
\begin{equation}
 \Ek = \sum_{\x} e^{-i\kk\cdot\x} \Ex\,,
\end{equation}
the energy correlation matrix in Fourier space is
\begin{align}
 \overline{\Ek\Ekp} &= \sum_{\x,\y} e^{-i\kk\cdot\x-i\kp\cdot\y}\,\overline{\Ex\Ey} =\\
                    &= N \sum_{\x,\y} e^{-i\kk\cdot\x-i\kp\cdot\y} q(\x-\y)^{\alpha N} =\,.
\end{align}
We can now define $\vec u=(\x+\y)/2$ and $\vec v=(\x-\y)/2$, so that
\begin{equation}
 = N 2^N \left[2^{-N} \sum_{\vec v} e^{i\vec v\cdot(\kp-\kk)} \right] \sum_{\vec u}e^{-i(\kp+\kk)\cdot\vec u} q(2u)^{\alpha N}
\end{equation}
where the term in square brackets is a representation of the Kronecker delta, $\delta_{\kk\kp}$,
so there is no correlation for any $\kk\neq\kp$.

Consequently, the variance can be written as 
\begin{equation}
 \overline{\Ek E(-\vec k)} = N2^N \sum_{\vec z} e^{-i\vec z\cdot\kk}q(\vec z)^{\alpha N}\,,
\end{equation}
which can be simplified to a form that is easily implemented numerically,
\begin{equation}
 \overline{E(\vec n) E(-\vec n)} =N2^N \sum_{\vec z} (-1)^{\sum_iz_in_i} \left(\frac1N\sum_{i=1}^N(-1)^{z_i}\right)^{\alpha N}\,.
\end{equation}

\section{Details of the calculations} \label{app:prob}
In this section we calculate explicitly some of the relations written in the main paper.
\subsection{Order of a configuration} \label{app:order}
The order of a configuration $\vec x$ is the number of neighboring configurations with energy $E^\mathrm{(neigh)}<E$.
Since each configuration has $N$ neighbors, the probability of a configuration of energy $E$ having order $k$
reduces to the probability that it have exactly $k$ neighbors with lower energy, and $N-k$
neighbors with $E^\mathrm{(neigh)}>E$. 

This amounts to taking the joint distribution of $N+1$ energies: the one of state $\vec x$, and its $N$ neighbors.
Calling $(E_0,\ldots,E_N)$ the vector representing these $N+1$ energies, 
the joint distribution can be written
as
\begin{align}
 P\big((E_0, &\ldots,E_N)\big) = \\\notag
             &\sqrt{\frac{\det(\mathcal{Q})}{(2\pi)^{N+1}}} \exp\left[(E_0,\ldots,E_N) \mathcal{Q}^{-1}\begin{pmatrix}E_0\\.\\.\\.\\E_N\end{pmatrix}\right]\,.
\end{align}
$\mathcal{Q}$ has a diagonal band structure, and its inverse $\mathcal{Q}^{-1}$ is tridiagonal \cite{meurant:92}.
As a consequence, its only non-zero non-diagonal elements are those relating each neighbor to $\vec x$.
This means that once the energy $E\equiv E_0$ of the configuration $\vec x$ is fixed, all the neighbors are mutually independent.

Therefore, the probability of a state being of order $k$ takes the binomial form
\begin{equation}
 P_k(E) = \binom{N}{k}\, 
          P\left(E^\mathrm{(neigh)}>E\right)^{N-k}\,
          P\left(E^\mathrm{(neigh)}<E\right)^k\,,
\end{equation}
which through Eq.~\eqref{eq:pe-cond} can be rewritten as
\begin{widetext}
\begin{equation}\label{eq-app:pe-sad1}
 P_k(E) = \binom{N}{k}
 \left(\frac{1}{\sqrt{2\pi N(1-\rho^2)}} \int_{E}^{\infty} e^{-\frac{(E'-\rho E)^2}{2N(1-\rho^2)}}dE' \right)^{N-k}
 \left(\frac{1}{\sqrt{2\pi N(1-\rho^2)}} \int^{E}_{-\infty}e^{-\frac{(E'-\rho E)^2}{2N(1-\rho^2)}}dE' \right)^k \,.
 \end{equation} 
\end{widetext}
Through a variable change, the first integral can be rewritten as $\frac12\erfc[B(E)]$, with 
\begin{equation}\label{eq-app:B}
 B=-E\sqrt{\frac{(1-\rho)}{2N(1+\rho)}}\,,
\end{equation}
whereas the second one is equal to $\frac12\erfc[-B(E)]=1-\frac12\erfc[B(E)]$. Consequently,
\begin{align}
\label{eq-app:pe-sad2}
 P_k(E) &= \binom{N}{k} \left(   \frac{\erfc \left[B(E)\right]}{2} \right)^{k}       
                         \left(1-\frac{\erfc \left[B(E)\right]}{2} \right)^{N-k} \,.
\end{align} 

\subsection{Complexity} \label{app:complexity}
We now calculate the intensive (i.e. for $E\sim-N$) complexity, which is defined as
\begin{equation}\label{eq-app:complexity-def}
 \Sigma_k = \log\left(2^N P(E) P_k(E)\right)\,,
\end{equation}
where $P(E\frac{1}{\sqrt{2\pi N}}e^{-\frac{E^2}{2N}}$ is the distribution of the energies defining the model.

We are interested in finite $k$, with diverging $N$. In this limit, the binomial coefficient reduces to $N^k$.
Also, since we focus on intensive energies, the term $B(E)$ in Eq.~\ref{eq-app:pe-sad2} is large,
and one can make an asymptotic expansion of the complementary error function.
To first order, for large $x$, $\erfc(x) \simeq \frac{e^{-x^2}}{\sqrt{\pi}x}$, so, keeping only the
dominant order, one has
\begin{align}
 \Sigma_k(E) &= N\log(2) - \frac{E^2}{2N} - kB^2 =\\
             &= N\log(2) - \frac{E^2}{2N}\left( 1- k\frac{1-\rho}{1+\rho}\right)\,.
\end{align}
\bigskip

\subsection{Threshold Energy} \label{app:eth}
A configuration at $\Eth$ typically has its lowest neighbor at its same energy: 
\begin{equation}\label{eq-app:emineth}
 E_\mathrm{min}^\mathrm{(neigh)}(E)=\Eth\,.
\end{equation}
In fact, this means that if $E<\Eth$, a configuration is typically a minimum, whereas
for $E>\Eth$ it typically is not.

As shown in Eq.~\ref{eq:pe-cond}, given a configuration at energy $E$, its neighbors' energies
follow a Gaussian distribution centered in $\rho E$, with variance $\sigma^2=N(1-\rho^2)$.

The typical minimum of $N$ Gaussians of variance $\sigma^2$ and mean $\mu$ is positioned at 
$\mu-\sqrt{2\sigma^2\log(N)}$ 
\footnote{Finite-size corrections to this result are of order $\log(\log(N))$, see e.g. \cite{gumbel:58}}. 
Therefore, the lowest neighbor
of a configuration with energy $E$ has energy
\begin{equation}
E_\mathrm{min}^\mathrm{(neigh)}(E)=\rho E-\sqrt{2(1-\rho^2)N\log(N)}\,. 
\end{equation}
By solving condition \eqref{eq-app:emineth} for $\Eth$, one obtains
\begin{equation}
 \Eth = -\sqrt{2\frac{1+\rho}{1-\rho}N\log(N)}\,.
\end{equation}

\bibliography{marco}

\end{document}